\documentclass[12pt]{article}
\usepackage{amsmath}
\usepackage{amsfonts}
\usepackage{feynmf}
\newcommand{\bean}{\begin{eqnarray*}}
\newcommand{\eean}{\end{eqnarray*}}
\newcommand{\ed}{\end{document}}

\newcommand{\be}{\begin{equation}}
\newcommand{\ee}{\end{equation}}
\newcommand{\barr}{\begin{array}}
\newcommand{\earr}{\end{array}}
\newcommand{\bea}{\begin{eqnarray}}
\newcommand{\eea}{\end{eqnarray}}

\begin{document}
\title{Perturbation theory for nonlinear equations}
\author{A.V.Bratchikov
 \\ Kuban
State Technological University,\\2 Moskovskaya Street, Krasnodar,
  350072, Russia
\\
}
\date{} \maketitle
\begin{abstract}
For a wide class of nonlinear equations a perturbative solution is constructed. This class includes equations of motion of field theories. The solution possesses a graphical representation in terms of diagrams. To illustrate the formalism we consider the Yang-Mills field equations.
\end{abstract}



\section{ Introduction}
Let ${ V}$ and ${W}$ be  vector spaces, 
and let 
\bea \label {qu}
F(u)= 0
\eea
be an equation, where $F:  { V} \to { W}$ is a  given function, and  $u\in {V}$ is an unknown vector.
For a given $u_0\in V$ one can expand  the equation in powers of $v=u-u_{0},$ 
\bea  \label{A}
F_0+ 
\sum_{n=1}^\infty 
F_n(v)=0.
\eea
Here $F_0=F(u_0),$ and for $n\geq 1$
\bea  
\label{b}
F_n(v)=\left.\frac 1 {n!}
\frac {d^n F(u_{0}+\xi v)} 
{d^n\xi}\right|_{\xi=0}.
\eea

Let $v_{0}$ be the general solution of the free equation
\bea \label{hom}
F_1(v)=0,
\eea
and let ${\cal E} (f)$ be the specific solution of the inhomogeneous equation 
\bea \label{un}
F_1(v)=-f,\quad f\in { W},
\eea
which is linear in $f,$ ${\cal E} (af)=a{\cal E}(f),a\in {\mathbb R}.$
Equation  (\ref {A}) then becomes 
\bea \label{bas}
v =w_0+\sum_{n=2}^\infty p_n(v),
\eea
where $w_0= v_0+ p_0,$  
$p_n
={\cal E} ( 
F_n
).$

Equation (\ref {qu}) is rather general. It appears in many problems. 
In particular, equations of motion of 
field 
theories are written in the form (\ref {qu}). For most of nonlinear equations in mathematical physics  
solutions of equations (\ref{hom}) and (\ref {un}) are well known \cite {CG}.

The aim of the present paper is to construct a solution of equation (\ref {bas}).The solution possesses a graphical representation in terms of diagrams.To illustrate the formalism  we consider the
pure Yang-Mills field equations in the Lorentz gauge.A solution of the equations of motion in spinor electrodynamics was found in ref.\cite {B}.  

The paper is organized as follows.  In the next section we construct a solution of equation (\ref{bas}) and describe its graphical representation. 
In Sec.3 we consider the Yang-Mills field equations.

\section {A solution of nonlinear equations}
We shall need 
functions $$\langle \ldots.  \rangle_n:
{V }^n 
 \to V,\quad n=2,3,\ldots,$$
defined for $v_1,\ldots,v_n\in V$ by 
\bea \label {orsk}
\langle v_1,\ldots,v_n \rangle_n =\sum_{r=1}^{n}(-1)^{n-r}
\sum_{i_1< \ldots <i_r} p_n( v_{i_1}+\ldots+ v_{i_r}).
\eea 
One can show \cite {L} that $\langle v_1,\ldots,v_n \rangle_n$ is an $n-$ linear symmetric function.

From (\ref{orsk}) it follows
\bean \label {or}
\langle v,\ldots,v \rangle_n  = {n!} p_n(v), 
\eean 
Then equation (\ref{bas}) takes the form 
\bea \label {equv}
v= w_0+\sum_{n=2}^\infty \frac 1 {n!} \langle v,\ldots,v \rangle_n.
\eea

For $I=(i_1,\ldots, i_n)$ we denote $v_I=(v_{i_1},\ldots,v_{i_n}).$
To solve equation (\ref {equv}) we introduce a family of functions $$\langle \ldots.  \rangle:
{V }^m 
 \to V,\qquad m=2,3,\ldots,$$
recursively defined by 
\bea \label {ord}
\langle v_1,\ldots,v_m \rangle =\sum_{n=2}^{\infty} \frac 1 {n!} 
\sum_{I_1\cup \ldots \cup I_n=(1\ldots m)} \langle \langle v_{I_1}\rangle,
\langle v_{I_2}\rangle,\ldots, \langle v_{I_n}\rangle 
 \rangle_n,
\eea 
where $I_i,i=1,\ldots,n,$ is an increasing multi-index \footnote{The multi-index $I=(i_1,\ldots ,i_n)$ is said to be increasing if $i_1< \ldots < i_n.$ }, $I_i\cap I_j=\emptyset,$ $\langle v \rangle=v,$ and for 
$I=(i_1,\ldots, i_n)$ $\langle v_I\rangle =\langle v_{i_1},\ldots,v_{i_n}\rangle.$
It is easy to prove by induction that $ \langle v_1,\ldots ,v_m \rangle$ is an $m-$ linear symmetric function.

For $m=2$ and $m=3$ we have
\bean \label {or}
\langle v_1,v_2 \rangle = \langle v_{1},v_{2}\rangle_2,
\eean 
\bean \label {or}
\langle v_1,v_2,v_{3}\rangle=
\langle \langle v_{1},v_{2}\rangle_2, v_3 \rangle_{2}+\langle \langle v_{1},v_{3}\rangle_2, v_2 \rangle_{2}+\langle \langle v_2, v_3 \rangle_2, v_{1}\rangle_2+\langle v_1, v_2,v_3 \rangle_3. 
\eean

Let $P^m_{i_1\ldots i_n}: V^m\to V^{m-n+1}, m\geq 2, 1\leq i_1< \ldots < i_n\leq m,$ be defined by
\bean \label {or}
P^m_{i_1\ldots i_n}( v_1,\ldots ,v_m ) = 
( \langle v_{i_1},\ldots {v}_{i_n}\rangle_n ,v_1,\ldots,\widehat{v}_{i_1},\ldots,\widehat{v}_{i_n},\ldots,v_{m} ),
\eean 
where $\widehat{v}$ means that ${v}$ is omitted.
If $v\in V$ is given by  
\bea \label {v} v = P^{n_s}_{I_s}\ldots P^{m-n_1+1}_
{I_2}
P^{m}_{I_1}
(v_1,\ldots ,v_m )
\eea
for some $I_1= (i_1^1,\ldots,  i^1_{n_1}),
I_2=(i^2_1,\ldots, i^2_{n_2}),
\ldots,I_s=(i^s_1,\ldots, i^s_{n_s}), n_1+\ldots+n_s-s+1=m,
$ we say that  $v$ is a descendant of $(v_1,\ldots ,v_m ).$

The functions  $\langle v_1,v_2\rangle$ and $\langle v_1,v_2,v_{3}\rangle$ are given by the  sums of all the descendants of their arguments.
Assume that $\langle v_1,\ldots,v_k \rangle, k<m, $ is given by the sum of all the descendants of  $(v_1,\ldots,v_k ).$
Each descendant of  $(v_1,\ldots ,v_m )$ can be written as
\bea \label{dec}\langle d_1(v_{I_{i_1}}),
d_2(v_{I_{i_2}}),\ldots , d_n(v_{I_{i_n}})\rangle_n,\eea for some $n\geq 2$ and ${I_{i_1}\cup \ldots \cup I_{i_n}=(1,\ldots, m)},$
where $I_{i_k}$ is an increasing multi-index, $d_k(v_{I_{i_k}})$ is a descendant of  $v_{I_{i_k}},k=1,\ldots, n.$ 
It is easy to show that for $k\neq l$ $I_{i_k}\cap I_{i_l}=\emptyset.$
Then summing all the different functions (\ref {dec}) we get the right-hand side of (\ref{ord}).
Thus, we have proved that $\langle v_1,\ldots,v_m \rangle$  is given by the sum of all the descendants of  $(v_1,\ldots ,v_m ).$

Each  descendant can be represented by a diagram. In this diagram an element of  $ V$ is represented by the line segment \rule[3pt]{30pt}{0.5pt}. A product $\langle v_1,\ldots,v_{n}\rangle_n$ is represented by the vertex joining the line segments for     $ v_1,\ldots,{v}_{n}$ and $\langle v_1,\ldots,v_n\rangle_n.$ The general rule should be clear from Figure~1.Here we show the diagram for 
$$P^m_{ij}(v_1,\ldots,v_m)=( \langle v_{i},{v}_{j}\rangle_2 ,v_1,\ldots,\widehat{v}_{i},\ldots,\widehat{v}_{j},\ldots,v_{m} ).$$
The points labeled  by $1,\ldots, m$ represent the ends of the lines $v_1,\ldots ,v_m.$  Using this prescription, one can consecutively draw the diagrams for 
$P^{n_1}_{I_1}
(v_1,\ldots ,v_m ),\\P^{n_2}_
{I_2}
P^{n_1}_{I_1}
(v_1,\ldots ,v_m ),\ldots , v$ (\ref {v}). 
The diagram for $v$ has $m+1$ external lines.
The auxiliary  points  $1,\ldots, m$ are removed.

\bigskip
\begin{center}
\begin{fmffile}{graph571}
{
\begin{fmfgraph*}(90,70)
\fmfpen {thin}
\fmflabel{$\langle v_i,v_j\rangle_2 $}{o1}
\fmfleft{i1,i2,i3,i4,i5,i6,i7}
\fmfdot{i1,i2,i4,i6,i7}
\fmflabel{$1$}{i1}
\fmflabel{$i$}{i3}
\fmflabel{$j$}{i5}
\fmflabel{$m$}{i7}
\fmf{plain,label=$v_i$,label.side=right}{i3,v1}
\fmf{plain,label=$v_j$,label.side=left}{i5,v1}
\fmf{plain,$\langle v_i,v_j\rangle_2 $,label.side=right}{v1,o1}
\fmfright{o1}
\fmfdot{i1,i2,i3,i4,i5,i6,i7}
\end{fmfgraph*}} 
\end{fmffile}
\end{center}
\bigskip
\begin{center}
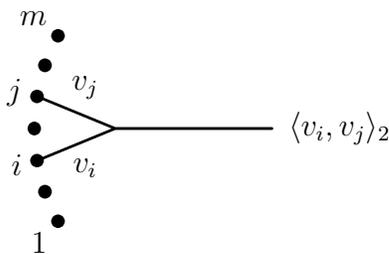
Figure 1. Diagram for $P^m_{ij}(v_1,\ldots,v_m).$
\end{center}
\bigskip


For $v_1=\ldots =v_m= w_0$ equation (\ref {ord}) reads
\bea \label {orss}
\langle w_0^m \rangle =\sum_{n=2}^\infty\frac 1 {n!} \sum_{s_1+\ldots +s_n=m}\frac {m!} {s_1!\ldots s_n!} 
\langle \langle w_{0}^{s_1} \rangle,\ldots,
\langle w_0^{s_n}\rangle \rangle_n,
\eea 
where $\langle w_0^r \rangle= \langle \underbrace{w_0,\ldots ,w_0}_{r 
}\rangle.$

We find that a solution of equation (\ref {equv}) is given by \bea \label {orro}
v= \langle e^{w_0} \rangle, 
\eea
where
\bean \label {}
\langle e^{w_0} \rangle = \sum_{k=0}^\infty \frac {1} {k!}\langle w_0^k \rangle, \quad \langle w_0^0 \rangle=0.
\eean
Indeed, substituting (\ref {orro}) in (\ref {equv}), we get 
\bean \label {or}
\sum_{m=2}^\infty \frac 1 {m!} \langle w_0^m \rangle =\sum_{m=2}^\infty \sum_{n=2}^\infty \frac 1 {n!} \sum_{s_1+\ldots +s_n=m}\frac {1} {s_1!\ldots s_n!} 
\langle \langle w_{0}^{s_1} \rangle,\ldots,
\langle w_0^{s_n}\rangle \rangle_n.
\eean 
To conclude the proof, it remains to use  (\ref {orss}).

Let 
\bean 
\xi= \xi_0+\sum_{n=2}^\infty \frac {\epsilon_n} {n!} \xi^n
\eean
be an  equation,
where $\xi,\xi_0\in \mathbb R,$ $\epsilon_n$ is defined as $0$ if $\langle v,\ldots,v \rangle_n=0,$ and otherwise $\epsilon_n=1.$ Let $g:\mathbb R \to \mathbb R$ be defined by $$g(\xi)=\xi-\sum_{n=2}^\infty \frac {\epsilon_n} {n!} \xi^n,$$
and let $g^{-1}(\xi)=\sum_{m=1}^\infty a_m \xi^m$ be the inverse function of $g.$ Then $$\langle e^{\xi} \rangle=\sum_{m=1}^\infty a_m \xi^m,$$ where $\langle \xi,\ldots,\xi \rangle_n=\xi^n,$ and therefore 
the number of the descendants of $(v_1,\ldots,v_m)$ is given by $m!a_m.$  

\section {The Yang-Mills field equations}
Let $x=(x^0,x^1,x^2,x^3)
$ be space-time coordinates, $\eta^{\mu\nu}=diag (1,-1,-1,-1)$ the metric tensor  and $\Box=\partial^\mu \partial _\mu.$
The pure Yang-Mills equations in the Lorentz gauge
$\partial^\mu A_\mu=0$ read \cite {FS}
\bea  \label{Y}
\Box A_\nu+[A^\mu,(\partial_\nu A_\mu -\partial_\mu  A_\nu +[ A_\mu, A_\nu])]-\partial^\mu[ A_\mu, A_\nu]=0,
\eea
where $A_\mu(x)$ is a non-Abelian gauge field.
For $u_0=0,F_0=0,v=A_\nu dx^\nu,$
\begin {gather*}
F_1(v)=\Box A_\nu dx^\nu,
\quad    
F_2(v)=(\partial^\mu[ A_\mu, A_\nu]-[A^\mu,(\partial_\nu A_\mu -\partial_\mu  A_\nu)])dx^\nu,
\\
F_3(v)=-[A^\mu,[ A_\mu, A_\nu]]dx^\nu
\end{gather*}
and $F_n(v)=0, n\geq 4,$  equations (\ref{Y}) are identified with (\ref {A}).  

The general solution $v_0=A_{0\nu} dx^\nu$ of the free equation
\bean \label{U}
\Box v=0
\eean
is given by 
\bean
v_0= \frac 1 {4\pi}M(\psi)+\frac 1 {4\pi}\frac \partial {\partial t}(tM(\varphi)),
\eean
where 
\bean
M(\mu)= \int_{S} 
\mu( x^1+t\xi^1, x^2+t\xi^2, x^3+t\xi^3) d \sigma_\xi, 
\eean
$\xi^1, \xi^2$ and $\xi^3$ are coordinates on the unit sphere $S$, $\sigma_\xi$ is the area element on $S,$  $t=x^0,$
$$v_0(0,x^1,x^2,x^3)=\varphi(x^1,x^2,x^3),\quad \left.\frac {\partial v_0(t,x^1, x^2,x^3)}{\partial t}\right|_{t=0}=\psi(x^1, x^2,x^3).$$
A specific solution of the inhomogeneous equation
\bean \label{U}
\Box v=-f
\eean
reads \cite {CG}
\bean
{\cal E}(f)=-\frac 1 {4\pi}\int^t_0\tau d\tau \int_{S} 
f(t-\tau, x^1+\tau \xi^1, x^2+\tau \xi^2, x^3+\tau \xi^3) d \sigma_\xi. 
\eean 

We can rewrite equation  (\ref {Y}) in the form (\ref {equv})
\bea \label {equvv}
A= A_0+\sum_{n=2}^\infty \frac 1 {n!} \langle A,\ldots,A \rangle_n,
\eea
where $A=A_{\nu} dx^\nu,$$A_0=A_{0\nu} dx^\nu.$
Combining (\ref {equvv}),(\ref {equv}) and (\ref {orro}), we get   
\bean \label {ororo}
A= \langle e^{A_0} \rangle. 
\eean
   
For example, the $O(A_0^3)$ contribution in $A$ is given by  
\bean
\frac 1 6\langle 
A_0,
A_0, 
A_0\rangle
=\frac 1 2
\langle\langle A_0, A_0\rangle_2, A_0\rangle_2+\frac 1 6\langle 
A_0,
A_0, 
A_0\rangle_3.
 \eean
The diagram for $ \langle\langle A_0, A_0\rangle_2, A_0\rangle_2 $ is depicted in Figure 2. Here the Yang-Mills fields are represented by wavy lines.

\bigskip
\begin{center}
\begin{fmffile}{graph573}
{\begin{fmfgraph*}(90,70)
\fmfpen{thin}
\fmflabel{$\langle\langle A_0, A_0\rangle_2, A_0\rangle_2 $}{o1}
\fmfleft{i1,i2,i3
}
\fmflabel{$A_0$}{i1}
\fmflabel{$A_0$}{i2}
\fmflabel{$A_0$}{i3}
\fmf{photon}{i2,v1}
\fmf{photon}{i1,v1,v2,i3}
\fmf{photon}{v2,o1}
\fmfright{o1}
\end{fmfgraph*}}
\end{fmffile}
\end{center}
\bigskip
\begin{center}
Figure 2. Diagram for $\langle\langle A_0, A_0\rangle_2, A_0\rangle_2$
\end{center}
\bigskip


\end{document}